\documentstyle[emulateapj]{article}

\def\simlt{\mathrel{\hbox{\rlap{\hbox{\lower4pt\hbox{$\sim$}}}\hbox{$<$}}}}
\def\simgt{\mathrel{\hbox{\rlap{\hbox{\lower4pt\hbox{$\sim$}}}\hbox{$>$}}}}
\def\gsim{\;\rlap{\lower 2.5pt
\hbox{$\sim$}}\raise 1.5pt\hbox{$>$}\;}
\def\lsim{\;\rlap{\lower 2.5pt
   \hbox{$\sim$}}\raise 1.5pt\hbox{$<$}\;}

\def\spose#1{\hbox to 0pt{#1\hss}}
\def\lta{\mathrel{\spose{\lower 3pt\hbox{$\mathchar''218$}}
     \raise 2.0pt\hbox{$\mathchar''13C$}}}
\def\gta{\mathrel{\spose{\lower 3pt\hbox{$\mathchar''218$}}
     \raise 2.0pt\hbox{$\mathchar''13E$}}}
\newcommand{\beq}{\begin{equation}}
\newcommand{\eeq}{\end{equation}}
\newcommand{\HDblah}{HD~$\!$209458~$\!$b}
\newcommand{\LR}{$L_{D}$}

\newcommand{\Tatm}{$T_{\rm atm}$}


\lefthead{CHO ET AL.}
\righthead{CHANGING FACE OF HD 209458b}

\begin{document}

\title{Changing Face of the Extrasolar Giant Planet, \HDblah}

\author{James Y-K. Cho,\altaffilmark{1,5} Kristen
Menou,\altaffilmark{2,6,7} Bradley M. S. Hansen\altaffilmark{2,3,8} \&
Sara Seager\altaffilmark{4,5}}

\altaffiltext{1}{Spectral Sciences Inc., 
	99 South Bedford Street, \#7, Burlington, MA 01803, USA}
\altaffiltext{2}{Princeton University, Department of Astrophysical
	Sciences, Princeton, NJ 08540, USA}
\altaffiltext{3}{Division of Astronomy, UCLA, 8971 Math Sciences, 
	Los Angeles, CA 90095, USA} 
\altaffiltext{4}{School of Natural Sciences, Institute for Advanced 
	Study, 1 Einstein Drive, Princeton, NJ 08540, USA}

\altaffiltext{5}{Present address: Carnegie Institution of Washington, 
	Department of Terrestrial Magnetism, 5241 Broad Branch Rd. NW, 
	Washington, DC 20015, USA}
\altaffiltext{6}{Chandra Fellow}
\altaffiltext{7}{Present address: Department of Astronomy, 
	P.O. Box 3818, University of Virginia, Charlottesville, 
	VA 22903, USA}
\altaffiltext{8}{Hubble Fellow}

\vspace{\baselineskip}

\begin{abstract}
High-resolution atmospheric flow simulations of the tidally-locked
extrasolar giant planet, \HDblah, show large-scale spatio-temporal
variability.  This is in contrast to the simple, permanent day/night
(i.e., hot/cold) picture.  The planet's global circulation is
characterized by a polar vortex in motion around each pole and a
banded structure corresponding to $\sim$3 broad zonal (east-west)
jets.  For very strong jets, the circulation-induced temperature
difference between moving hot and cold regions can reach up to
$\sim$1000~K, suggesting that atmospheric variability could be
observed in the planet's spectral and photometric signatures.

\end{abstract}

{\it subject headings}: planetary systems -- planets and satellites:
general -- stars: atmospheres -- turbulence

\section{Introduction}

Roughly 100 gaseous giant planets are currently known to orbit nearby
sun-like stars\footnote[9]{see, e.g., {\tt
http://www.obspm.fr/encycl/encycl.html} and {\tt
http://exoplanets.org/almanacframe.html}}.  Many of those planets are
located at very small orbital distances from their parent stars, where
tidal forces are thought to maintain a rotation rate synchronous with
the orbit---thus producing permanent day and night sides on the
planet.  This situation presents a new r\'egime of atmospheric
circulation, not encountered in our solar system: slowly-rotating
giant planets, which are continuously exposed to intense stellar
heating on the same side.  Measuring the resulting temperature
structure on these planets is a major goal of current and future
observational programs.  In this {\it Letter}, we report on
high-resolution, fully-turbulent global simulations of the atmospheric
flow on \HDblah, presently the only close-in extrasolar giant planet
(CEGP) with a measured mass ($M_p$) and radius ($R_p$).

The parent star of \HDblah\ shows discernible brightness decrements
every 3.5~days, due to occultations by the planet as it transits
across the star.  This property has recently led to precise
measurements of $M_p$ and $R_p$ (Charbonneau et al. 2000; Henry et
al. 2000; Mazeh et al. 2000), confirming its giant nature.  It has
also allowed the detection of sodium absorption, providing the first
probe of the planet's atmosphere (Charbonneau et al. 2002).  According
to the standard planetary formation picture, \HDblah\ is expected to
have formed at a large distance ($>$~1~AU) from its parent star (see,
e.g., Boss 1996) and migrated inward (Goldreich \& Tremaine 1980; Lin
et al. 1996; Murray et al. 1998), quickly ($\simlt$~10 Myr) reaching
its present distance of only 0.046~AU from the star (see e.g., Burrows
et al. 2000).  There, it was forced by tidal effects to permanently
present the same face to its star (see, e.g., Goldreich \& Soter
1966)---as the Moon does to the Earth.  From this synchronization, the
rotation period of the CEGP is known (same as its orbital period of
3.5 days).  However, unlike our Moon with its insignificant
atmosphere, \HDblah\ is expected to possess vigorous meteorology and
associated horizontal transport of heat and chemical species, due to
the presence of a thin, stable (radiative) atmospheric region above
the convective interior (Guillot et al. 1996; Seager \& Sasselov
1998).

\section{Model}

We model the stable region of \HDblah's atmosphere as a shallow layer
of hydrostatically balanced, frictionless gas---moving under the
influence of gravitational and Coriolis accelerations.  The motion of
such a layer enveloping a planet is governed by the equivalent
barotropic formulation of the shallow-water equations on a rotating
sphere (Salby 1989):
\begin{eqnarray}
\frac{\partial {\bf v }}{\partial t}  +  {\bf v}\!\cdot\!\nabla\ {\bf
v } & = & -g\nabla h - f{\bf k}\times{\bf v} + {\cal F}_a,\\
\frac{\partial h}{\partial t}  +  {\bf v}\!\cdot\!\nabla\ h & = &
-{\cal K} h \nabla\!\cdot\!{\bf v} + {\cal F}_d,
\end{eqnarray}
where ${\bf v}\! =\! {\bf v}\,(\lambda,\varphi,t)$ is the horizontal
velocity with $\lambda$ and $\varphi$ the longitude and latitude,
respectively; $h\! =\! h\,(\lambda,\varphi,t)$ is the thickness of the
modeled layer, proportional to the temperature; $f\!  =\! 2\Omega
\sin\varphi$ is the Coriolis parameter, where $\Omega$ is the rotation
rate of the planet; ${\bf k}$ is the local unit vector normal to the
planetary surface; $g$ is the gravitational acceleration; ${\cal
F}_a\!  =\!  -g\nabla
\eta(1-\exp\{-t/\tau_a\})\cos(\varphi)\cos(\lambda)$ and ${\cal F}_d\!
=\!  -(h-h_{E})/\tau_{d}$ are, respectively, the adiabatic and
diabatic thermal forcing, representing uneven hemispheric heating and
cooling due to synchronization; $\tau_a$ and $\tau_d$ are
characteristic $e$-folding times; $h_E = H +
\eta\cos(\varphi)\cos(\lambda)$ is the equilibrium thickness; $\eta$,
and $H$ are constants; and, ${\cal K} = R/c_p$, where $R$ is the
specific gas constant and $c_p$ is the specific heat at constant
pressure.

The nonlinear equations (1) and (2) describe circulations in which the
motion field is vertically aligned over one or more pressure scale
heights, $H_p$.  They are column-integrated representation of the
equations used in general circulation models and are solved
numerically using the highly-accurate pseudospectral algorithm
(Eliasen, Mechenhauer \& Rasmussen 1970; Orszag 1970).  Approximately
150 simulation runs with up to T341 (1024$\times$512 grid) resolution
have been performed to explore the full physical and numerical
parameter-space available for the \HDblah\ atmosphere.  The details of
the full exploration are described elsewhere (Cho et al. 2003).  Here,
we emphasize only the main robust features from the large number of
simulations---namely, the emergence of long timescale variability due
to moving polar vortices and ``thermal'' spots of high contrast from
the background temperature field.

High-resolution, shallow-layer barotropic models in spherical geometry
have been successfully used to model the atmospheric dynamics of the
solar system gaseous giant planets (Williams 1978; Cho \& Polvani
1996a), as well as stratospheric (stable region) phenomena on the
Earth (see, e.g., Juckes \& McIntyre 1987).  An alternate model of
observed cloud level circulation on Jupiter as a surface expression of
deep convective columns, oriented parallel to the planetary rotation
axis, also exists (Sun et al. 1993; Schubert \& Zhang 2000).  However,
only the shallow-layer model has thus far qualitatively reproduced the
essential features (bands, zonal winds, persistent spots, and
anticyclonic dominance) on all of the giant planets in our solar
system using only the observed values of physical parameters (Cho \&
Polvani 1996a).

For \HDblah, we adopt parameter values where known (Charbonneau et
al. 2000; Henry et al. 2000; Mazeh et al. 2000): ${\cal K}\! =\!
0.29$, $g\! =\!  8$~m~s$^{-2}$, $R_p\! =~10^8$~m, $\Omega\! =\!
2.1\!\times\!  10^{-5}$~s$^{-1}$, and $H\! =\!  2\!\times\!  10^6$~m.
The last (=~$\! H_p/{\cal K}$) represents the global mean layer
thickness, whose value corresponds to the equivalent blackbody
temperature (\Tatm~$\approx\!  1400$~K for a Bond albedo, $A\!  =\!
0.2$; the results presented here, however, are insensitive to the
precise choice of \Tatm\ and $A$).  The values for two required
parameters---$\eta$, the amplitude of hemispheric thermal forcing, and
$U$, the characteristic wind speed related to the global mean kinetic
energy---are not known and hence varied.  From run to run, $\eta$ is
varied from 0 to $0.4 H$, corresponding respectively to no heating and
maximum substellar-point heating in the absence of atmospheric motion;
$U$ is varied from 50~m~s$^{-1}$ to 1000~m~s$^{-1}$, roughly the
observed value for Jupiter (Ingersoll 1990) and the value expected
from a simple thermal wind balance (Showman \& Guillot 2002),
respectively.  Note that the latter value leads to a
nonlinearly-balanced wind field which can locally approach the sound
speed ($\sim$2700~m s$^{-1}$); on all observed atmospheres in our
solar system, $U$ does not exceed $\sim$400~m~s$^{-1}$ (Beatty \&
Chaiken 1990).  We have carefully studied the effects of $U$ on the
variability and thermal contrast and find that, while a variability in
the flow is a generic feature for \HDblah, a significant thermal
contrast generally requires a small $\eta$ or a large $U$.  In the
following, we give an example of the expected generic flow pattern and
then describe a typical flow-induced thermal variability which may be
observationally relevant.

\section{Results}

Fig.~1 shows an instantaneous atmospheric flow field on \HDblah\ at
dynamic equilibrium from one of our simulations with $U\! =\!
400$~m~s$^{-1}$ and $\eta$ corresponding to a maximum day/night
temperature difference of 180~K.  The flow exhibits considerable
complexity and variability: it is characterized by a broad equatorial
band containing thin, undulating filamentary structures at low and
middle latitudes and displaced, translating circumpolar vortices at
the rotation poles.  The large undulations at low latitudes, which can
roll up into small vortices, are due to high-amplitude Rossby
(planetary) waves.  In addition to being responsible for weather, the
planetary waves efficiently stir and mix tracers (e.g., clouds,
chemical species, potential vorticity) when they break, homogenizing
the region in which the breakings occur (Juckes \& McIntyre 1987).  In
contrast, polar vortices sequester air inside, their sharp boundaries
serving as robust barriers to material transport between their
interiors and the well-mixed interstitial regions.  The overall flow
is similar to that of Earth's winter stratosphere.

The complex flow of Fig.~1 corresponds to a global circulation pattern
of 3 zonal jets, shown in Fig.~2.  Jets are atmospheric flow
structures, which are obtained by longitudinally averaging the
eastward wind at each latitude.  The pattern in Fig.~2 naturally
evolves from an initially random flow without jets, with the basic
profile forming approximately at day 20.  The emergence of jets from a
random field is a generic feature of turbulent flows in a
differentially rotating shallow layer (Rhines 1975; Williams 1978; Cho
\& Polvani 1996b).  For the extreme values of $U$ considered with all
other parameters held fixed, 2 or 4 jets are also possible.  However,
once formed, the jets do not change qualitatively---even under applied
thermal forcing---for up to several hundred \HDblah\ days, the
duration of the simulation.

The presence of a low number of broad jets, along with polar vortices,
leads to a circulation pattern which is markedly different from that
of cloud-top Jupiter.  Although CEGPs are sometimes dubbed ``hot
Jupiters'', the circulation pattern is more like that of Uranus or
Neptune, or even Earth, than that of Jupiter or Saturn, which have
$\sim$10 narrow jets (Ingersoll 1990).  Dynamically, this is expected
given roughly the factors of ten greater \Tatm\ and ten slower
rotation rate of \HDblah ~compared to Jupiter.  The combined effect,
giving the flows in Figs.\ 1 and ~2, is characterized by the Rossby
deformation scale, \LR$=\!({\cal K}gH)^{1/2}/f$, which is to be
compared with $R_p$.  Generally, \LR/$R_p \ll 1$ is required for a
Jupiter-like multiple-jet circulation pattern without pronounced polar
vortices at the cloud-top.  For \HDblah, \LR/$R_p\!\sim\! 1$.

Jets and vortices/eddies can strongly influence the overall thermal
structure by transporting heat.  Since the state of the \HDblah's
cloud-top is not likely to be at rest and driven solely by stellar
heating, we delay the application of thermal forcing (in contrast to
Showman \& Guillot 2002) until the flow field self-organizes into the
pattern in Fig.~2.  Dynamically, this is a likely flow state of
\HDblah\ in the presence of weak or uniform heating (i.e., at {\it
pre}-synchronization).  The forcing is slowly increased to the
specified value of $\eta$ over a characteristic radiative equilibrium
time, $\tau_r\!  =\!  10$~days, appropriate for $\sim$1~bar pressure
level on \HDblah.  At this depth, $\tau_r\gg\tau_c$ (where $\tau_c\!
=\!  R_p/U$ is the characteristic circulation time), signifying a
large distance advection of airmass before it warms or cools
appreciably.  The results presented here do not change qualitatively
with the value of $\tau_r$ ($\sim$ 0--100 days), using either an
adiabatic (${\cal F}_a, \tau_a\!  =\!\tau_r$) or a diabatic (${\cal
F}_d, \tau_d\!  = \!\tau_r$) representation of thermal forcing.

At each point on the planet, the layer thickness is related to the
temperature field, $T(\lambda,\varphi,t)\! =\! hg/c_p$, via the
hypsometric relation (Holton 1992; Salby 1989).  In our simulations
with small $\eta$ or large $U$, the $T$ field exhibits a generic
feature which is in direct contrast to the simple permanent day/night
picture.  Fig.\ 3 shows the $T$ distribution (from the same run of
Figs.~1 and 2) at several different times.  Note that the warmest and
the coldest regions are not located at the sub-stellar and the
anti-stellar ``thermal poles'' (at the equator), respectively.
Instead, the temperature extrema are located near the rotation poles,
inside a coherent hot/cold spot pair at each pole.  Nor are the
extrema stationary: they revolve around the poles with a period of
$\sim$25 days in this case.  The movement leads to a distribution that
for a time has the temperature minimum actually on the ``hot'' day
side and the maximum on the ``cold'' night side.  In addition, the
temperature difference between the two extrema is large and strongly
asymmetric; here, the difference is $\sim$300~K, with the diffuse hot
region $\sim$70~K above and the sharp cold region $\sim$230~K below
\Tatm.  For the case with $U\!  =\!  1000$~m~s$^{-1}$, the difference
reaches $\sim$1000~K, with a temperature minimum of $\sim$800~K below
\Tatm\ at the core of the cold spots.

\section{Discussion}

The presence of high-contrast hot and cold spots on \HDblah\ induces
spatio-temporal variability, which may produce detectable fluctuations
in observational signatures.  Sensitive enough infrared flux
measurements (e.g., such as those possible with SIRTF) could reveal
variability in time during an orbit of the planet as different faces
are seen from Earth, as well as from orbit to orbit as the spots
(polar vortices) revolve about the rotation poles.  The lack of
detectable variability would thus point toward either an obscuring
uniform haze overlying the modeled region or an inefficient conversion
of stellar irradiation to atmospheric kinetic energy.  In the latter
case, the formed spots are weak (i.e., small temperature/thickness
perturbation) and thermal forcing which produces bulging in the
modeled layer of more than several percent overwhelms any temperature
variability due to atmospheric motion.  The equilibrium day/night
temperature is then robustly maintained.  

The extreme conditions inside the spots in our calculations suggest
several additional potential observables since absorption levels,
albedo, intrinsic thermal emission, and presence, type, or height of
clouds could all be different within the different spots.  The
spatially integrated spectrum, therefore, can be different from the
uniform planet case.  For example, enhanced levels of CH$_4$/CO
abundance inside the cold spots are possible, given the value of
\Tatm\ used (Seager et al. 2000).  Similarly, condensates, such as
MgSiO$_3$ (enstatite), may also be found inside the diffuse hot spots,
where temperatures may be high enough (Sudarsky et al. 2000; Seager et
al. 2000).  If the temperature at radiative equilibrium is actually
higher than assumed (e.g., $\sim$2000~K), the opposite situation may
occur---MgSiO$_3$ may be found in the cold spots.  In addition, by
sequestering chemically active species and periodically exposing them
to the stellar irradiation (as the spots revolve around the poles),
the spots could also affect atomic number densities and could be part
of the explanation for the recently observed low abundance of Na~I on
\HDblah\ (Charbonneau et al.  2002).

\section*{Acknowledgments}
J.Y-K.C. acknowledges the hospitality of the Institute for Advanced
Study, where the initial part of this work was completed, and
R. Levine for helpful discussions.  K.M. is supported by NASA under
Chandra Fellowship grant PF9-10006, awarded by the Smithsonian
Astrophysical Observatory for NASA under contract NAS8-39073.
B.M.S.H. is supported by the Hubble Fellowship grant HF-01120.01-99A,
awarded by the Space Telescope Science Institute, which is operated by
the Association of Universities for Research in Astronomy, Inc.\ for
NASA under contract NAS 5-26555.  S.S. is supported by the W.M. Keck
foundation.  We thank the referee, W.\ Hubbard, for comments.

\clearpage

\begin{figure}[p]
\caption[]{Two views of the dynamical flow tracer, potential vorticity
(Holton 1992), at day (=year) 55 from our T341 (1024$\times$512 grid
resolution) simulation of the atmospheric circulation of \HDblah: a)
orthographic projection centered at the anti-stellar point (AS) on the
night side and b) polar-stereographic projection centered at the north
pole (NP).  1 PVU = 4 $\!\times\! 10^{-27}$~s$^{-1}$m$^{1/{\cal K}}$.
The global flow is characterized by two circumpolar cyclonic (rotating
in the same direction as the planet---counter-clockwise in the figure)
vortices at high latitudes and high-amplitude planetary waves at low
latitudes.
\label{fig:vorticity}}
\end{figure}

\begin{figure}[p]
\plotone{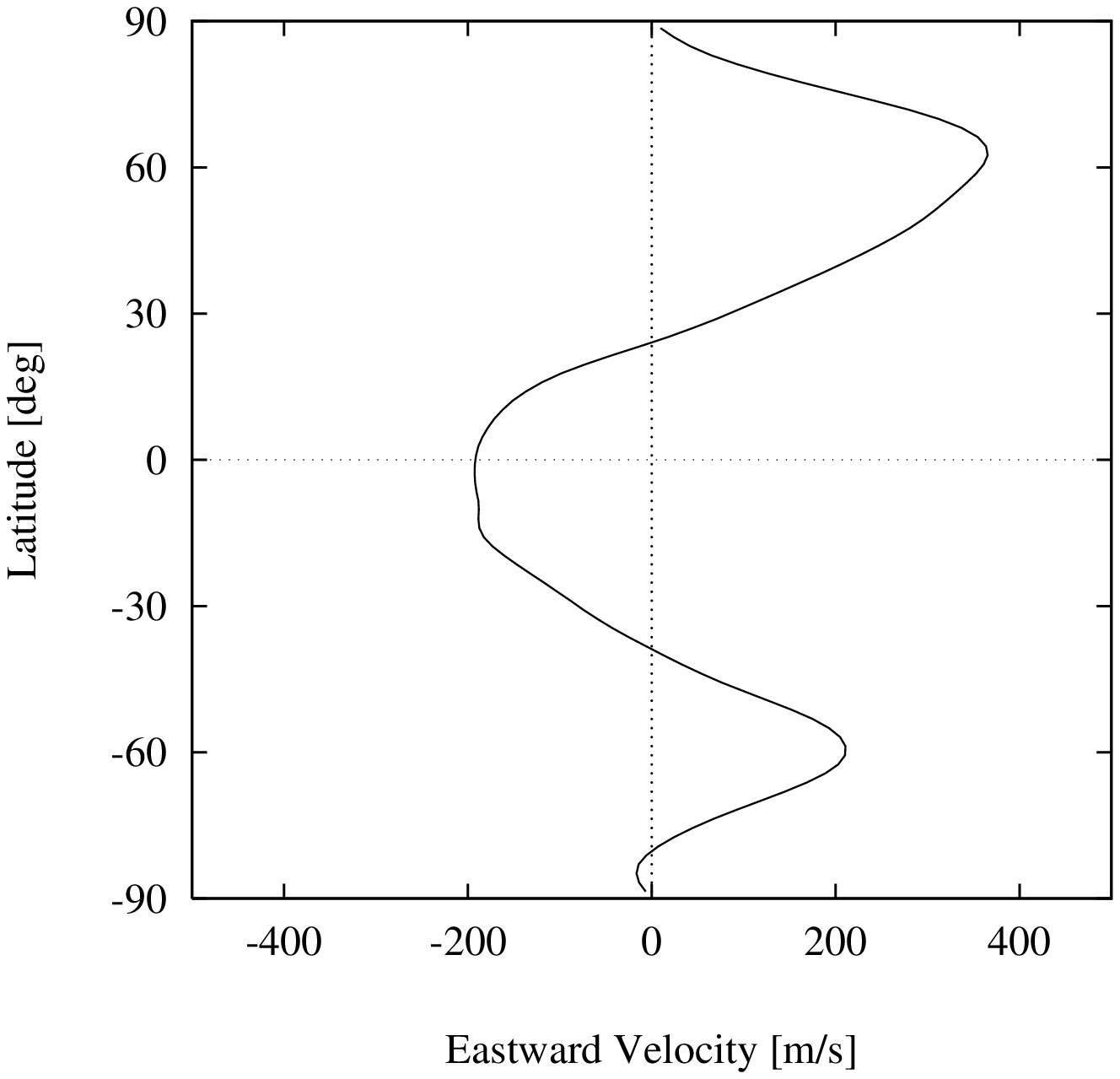}
\caption[]{Steady circulation profile from the flow of Fig.~1:
longitudinally-averaged zonal (eastward) wind velocity.  Zonal jets
emerge from a random initial condition due to conversion of
small-scale stirring/eddies to planetary waves in the atmosphere (as
seen in Fig.~1), a generic feature of shallow-layer turbulence under
differential (latitudinally-dependent) rotation $f$.  This basic
profile has emerged approximately at day~20 of the simulation and does
not change for the duration of the simulation (several hundred
\HDblah\ days).  On \HDblah, the jets are fewer, broader, and stronger
than those observed on Jupiter.  This is because \HDblah\ is hotter
and rotating more slowly than Jupiter.  Hemispheric thermal forcing is
applied on this robust flow.
\label{fig:wind}} 
\end{figure}

\begin{figure}[p]
\caption[]{Four successive night-side views of the temperature field
in orthographic projection (on \HDblah\ days indicated).  A persistent
``thermal dipole'', associated with the circumpolar vortices, revolves
around the pole and provides high thermal contrasts that may be
detectable in the planet's observational signatures.  In contrast to
the simple, permanent day/night picture, the hottest atmospheric
region is, at times, on the night side of the planet.  These spots are
long-lived, asymmetric (cold regions are more well-defined and have
larger amplitude), and can serve as large areas of atmosphere with
distinct chemistry and thermodynamics.  The global equilibrium
blackbody temperature of the atmosphere is taken to be 1400 K.
\label{fig:temperature}}
\end{figure}

\end{document}